\def\PRL{{Phys. Rev. Lett.\ }\/}
\def\PRB{{Phys. Rev. B\ }\/}

\def\be{\begin {equation}}
\def\ee{\end {equation}}
\def\ber{\begin {eqnarray}}
\def\eer{\end {eqnarray}}
\def\bers{\begin {eqnarray*}}
\def\eers{\end {eqnarray*}}

\makeatletter

\newcommand{\Rmnum}[1]{\expandafter\@slowromancap\romannumeral #1@}
\makeatother

\makeatletter
\newcommand*\env@matrix[1][*\c@MaxMatrixCols c]{%
  \hskip -\arraycolsep
  \let\@ifnextchar\new@ifnextchar
  \array{#1}}
\makeatother

\documentclass[aps,prb,showpacs,superscriptaddress,a4paper,twocolumn]{revtex4-1}
\usepackage{amsmath}
\usepackage{float}
\usepackage{color}

\usepackage[normalem]{ulem}
\usepackage{soul}
\usepackage{amssymb}

\usepackage{amsfonts}
\usepackage{amssymb}
\usepackage{graphicx}
\begin {document}

\title{Topological Nontrivial Phase in Hexagonal Antiperovskites A$_3$BiB (A=Ba,Sr\ ; B=P,N)}

\author{C. K. Barman}
\affiliation{Department of Physics, Indian Institute of Technology, Bombay, Powai, Mumbai 400 076, India}

\author{Chiranjit Mondal}
\affiliation{Discipline of Metallurgy Engineering and Materials Science, IIT Indore, Simrol, Indore 453552, India}
\author{Vijay Singh}
\affiliation{CEA, LITEN, 17 Rue des Martyrs, 38054 Grenoble, France}
\author{Biswarup Pathak}
\affiliation{Discipline of Metallurgy Engineering and Materials Science, IIT Indore, Simrol, Indore 453552, India}

\author{Aftab Alam}
\email{aftab@iitb.ac.in}
\affiliation{Department of Physics, Indian Institute of Technology, Bombay, Powai, Mumbai 400 076, India}
\date{\today}

\begin{abstract}
In this article, we predict the occurrence of topological non-trivial phase in hexagonal antiperovskite systems. By carefully investigating the evolution of band structure, we have studied the pressure induced topological phase transition in  Ba$_{3}$BiP, Ba$_{3}$BiN, and Sr$_{3}$BiN compounds using the hybrid functional calculations. The non-trivial topology has been verified by computing Dirac like surface dispersion and topological invariant {\it Z$_2$} index via parity analysis. The unconventional spin texture has been analyzed which guaranteed the absence of impurity induced backscattering on boundary of the sample while respecting the time reversal symmetry. Our simulation confirms the chemical and mechanical stability of all the three compounds. The present study introduces an important new class of hexagonal antiperovskite compounds in the topological regime and are believed to capture ample of attention both from theoretical as well as experimental front.
\end{abstract}
	
\maketitle

{\par}{\it \bf Introduction:}
The discovery of quantum materials with non-trivial band topology and robust surface states such as topological insulators (TIs),\cite{ZHCK2010,XQSCZ2011,JEMOORE,YANDO2013,TDAS2016}  topological semimetals\cite{SMYOUNG2012,ZWANG2012,LIU2014,JWRHIM2015,CKCHIU2016} have garnered significant attention among the scientific community. The non-trivial band topology of TIs is associated with exotic gapless surface/edge states at the interface of TI and ordinary insulator. The electron spins in these surface states are locked to their momenta. A time reversal symmetric (TRS) topological insulator are characterized by topological invariant {\it Z$_2$} index.\cite{KaneMele2005,FUKANE2007} In the TRS systems, the Kramer's degeneracy leads to a pair of states with opposite spin and momenta, which in turn suppress the impurity backscattering between these pair of states. The development to this area of research is mostly focused on the time reversal invariant TIs, also known as {\it Z$_2$}\cite{KaneMele2005,FUKANE2007} TI. Kane and Mele\cite{KaneMele2005} proposed to calculate the  {\it Z$_2$} index by counting the number of intersection between edge states and the Fermi level (E$_F$) between 0 to $\frac{\pi}{a}$ in the Brillouin zone. Soon after, several mathematical formulations were developed to generalize the {\it Z$_2$} topological invariants in three dimensions. Subsequently, Fu-Kane\cite{FUKANE2007} simplified the {\it Z$_2$} invariant calculation by computing the parity exchange\cite{BHZ2006,CYTEO2008,Hsieh-2008,YXIA2009,ZHANG2009} at the time reversal (TR) invariant momenta for inversion symmetric system. The  topological invariant, {\it Z$_2$} index indicates that the non-triviality of band topology is simply embedded in the band structure of an insulator, via band inversion. The band inversion mechanism can be tuned by a wide range of parameters including spin-orbit coupling,\cite{YXIA2009,ZHANG2009,NaBaBi2016, chiru2018} electron-electron,\cite{KONDO2010,SmB62013,YbB62016} electron-phonon interactions\cite{GARATE2013,KUSH2014,LLW2017} etc. Among these, the spin-orbit coupling (SOC) is one of the key factors, which is responsible to realize the band inversions or parity exchanges in most of the known TI genome. 

{\par} Owing to the SOC induced band inversion mechanism, a significant number of distinct topological classes have been theoretically predicted and experimentally verified. For example, the bismuth(Bi) based binary alloys Bi$_{1-x}$Sb$_x$\cite{CYTEO2008} and Bi$_2$Se$_3$\cite{Hsieh-2008,ZHANG2009,YXIA2009} were the first 3D topological insulators. Later, TI properties are also observed in other class of compounds like ternary thallium based chalcogenides,\cite{LIN_BANSIL2010,TAKAFUMI2010,CHEN2010,BYANEPL2010} Heusler alloys,\cite{HLIN2010,CHADOV2010,DXIAO2010,SAWAI2010,LIN2015,ZKLIU2016} actinides\cite{actinides} etc. 

{\par}In this article, we extend the understanding of topological insulating behavior in a new class of Bi$-$based antiperovskite systems. Antiperovskite systems are charge inverted materials to their perovskite counterpart. Similar to perovskite compounds, antiperovskite systems are extensively studied for their vast range of properties including unconventional superconductivity, \cite{cm1} unusual magnetism, \cite{cm2} and multiferroic orders.\cite{cm3} However, there are limited studies\cite{ANTIPEROVSKITE,PICKETT2018} where 3D TI has been realized. 
Although antiperovskite compounds are found to crystallize either in cubic\cite{MYC1992,NIEWA2004} or hexagonal\cite{NIEWA2004,NIEWA2013}  struc-
tures depending on their constituent elements, most of the theoretical studies in the past\cite{ANTIPEROVSKITE,PICKETT2018} are based on cubic phase due to its simplicity even if the reported experimental phase is hexagonal. In addition, most of the {\it ab-initio} studies utilize the standard generalized gradient approximation (GGA), which is know to be inappropriate for TIs (predicting band inversion and narrow gaps). It is thus extremely important to showcase how the prediction changes as one migrate from inaccurate GGA to a more accurate (say) hybrid functional calculation. This is one of the endeavor of the present manuscript.

In spite of a few studies on cubic antiperovskites, the investigation of TI properties in their hexagonal phase is still lacking. The purpose of our present study is to dig more into the family of hexagonal antiperovskites to realize the TI properties. We have simulated the cubic phase as well, and found it relatively less stable (see supplement\cite{supp}). Figure \ref{fig1} shows the crystal structure (space group of P6$_{3}$/mmc(\#194)) and Brillouin zone of bulk and (001) surface of hexagonal antiperovskite compounds.\cite{NIEWA2004,NIEWA2013} Using the first principle calculations based on  hybrid functional exchange correlation\cite{HSE2003}, we found three systems (Ba$_3$BiP, Ba$_3$BiN and Sr$_3$BiN) in their hexagonal phase which show promising topological insulating property in the family of antiperovskites. The topological insulating/semi-metallic phase is realized under pressure (both compression as well as expansion). Unlike conventional TIs, some of these compounds show band inversion between p- and d-orbitals (as oppose to s- and p-orbitals). This occurs due to the unique nature of hybridization between the neighboring atoms, which controls the movement of bands differently under strain. The non-trivial topology of bands is reconfirmed by calculating Dirac like surface states, spin texture and topological invariant Z$_2$ index (via parity analysis). Moreover, to get a better insight into the non-trivial band topology around $\Gamma$-point, we have illustrated our data using a low-energy effective Hamiltonian.

\par {\it \bf Computational Details}:\
First principle calculations were performed using Vienna Ab-initio Simulation Package (VASP)\cite{GKRESSE1993,JOUBERT1999} based on density functional theory (DFT). Projector Augmented Wave (PAW)\cite{PEBLOCHL1994} basis set was used with an energy cut off 500 eV. We adopt generalized gradient approximation by Perdew-Burke-Ernzerhof (PBE)\cite{JOUBERT1999} to describe the exchange and correlation. A more accurate estimation of band gap and band inversion strength was further checked by hybrid Heyd-Scuseria-Ernzerhof (HSE06)\cite{HSE2003}  exchange-correlation functional including SOC. Total energy (force) was converged upto 10$^{-5}$ eV/cell (0.01 eV/\AA). Brillouin zone integrations were performed using 7$\times$7$\times$9 $\Gamma$-centered k-mesh. The maximally localized wannier functions (MLWF)\cite{MARZARI1997,SOUZA2001,VANDERBIT} were used to construct tight-binding model to reproduce the band structure including SOC effects. Further iterative Green's function\cite{QuanSheng,DHLEE_I,DHLEE_II,SANCHO1985} scheme was employed to calculate surface spectral function. 

\begin{figure}[t]
\centering
\includegraphics[width=\linewidth]{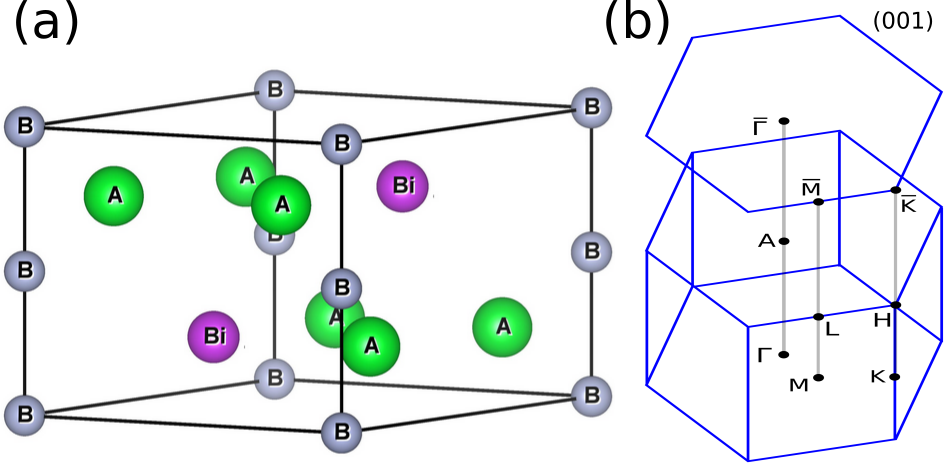}
\caption { (Color online) (a) Crystal structure (\# P6$_{3}$/mmc) and (b) Brillouin zone of bulk and (001) surface of hexagonal antiperovskites A$_3$BiB. A$=$Ba or Sr, and B$=$P or N. }
\label{fig1}
\end{figure}


{\par}{\it \bf Results and Discussion:}
Non-trivial band ordering in any TI can be predicted by band inversion of bulk states near the Fermi level. The band inversion has been realized in various materials by tuning its intrinsic properties, such as spin-orbit coupling, band hybridization by doping or alloying. Several families\citep{chiru2018,AAlam2013,YAN2014,KONG2011,TARAKANE2012,PDZIAWA2012} show doping induced topological phase transition (TPT) by means of band inversion. The pressure induced band inversion is another promising way to achieve TPT. Pressure induced TPT is becoming a popular avenue to find new TIs because it does not require any unwanted dopants or inhomogeneity of doping in the pristine system. So far, the pressure induced TPT has been achieved in a wide variety of systems.\citep{NaBaBi2016,uni2011,Ahuja2011,YANSUN2011,CKBAA2018} 

\begin{table}[t]
\begin{ruledtabular}
\caption{Relaxed lattice parameters (in \r{A}) of A$_3$BiB (A=Ba or Sr; B=P or N). Values within parenthesis are experimental values.\cite{NIEWA2004}   }
\label{table-1} 
\begin{tabular}{c c c}
System & a=b &  c \\
\hline
Ba$_3$BiP & 8.18 & 7.44 \\ 
Ba$_3$BiN & 7.76 (7.61) & 6.76 (6.68) \\ 
Sr$_3$BiN & 7.36 & 6.39 \\ 
\end{tabular} 
\end{ruledtabular}
\end{table}

\begin{figure}[b]
\centering
\includegraphics[width=\linewidth]{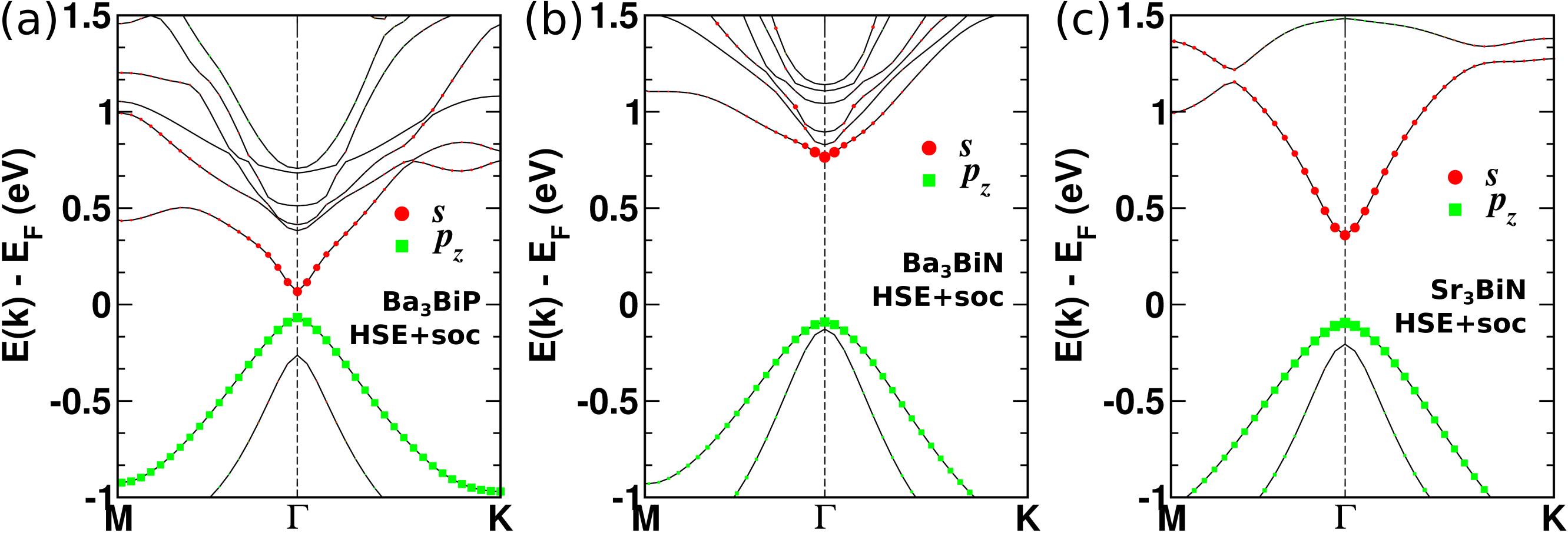}
\caption { (Color online) Bulk band structures of (a) Ba$_{3}$BiP, (b) Ba$_{3}$BiN  and (c) Sr$_{3}$BiN using HSE06 functional including SOC effect. Sizes of red and green dots represent weightage contribution of Bi(Ba,Sr){\it -s} and N(P){\it -p$_{z}$} orbitals. }
\label{fig2}
\end{figure}

\begin{figure}[t]
\centering
\includegraphics[width=\linewidth]{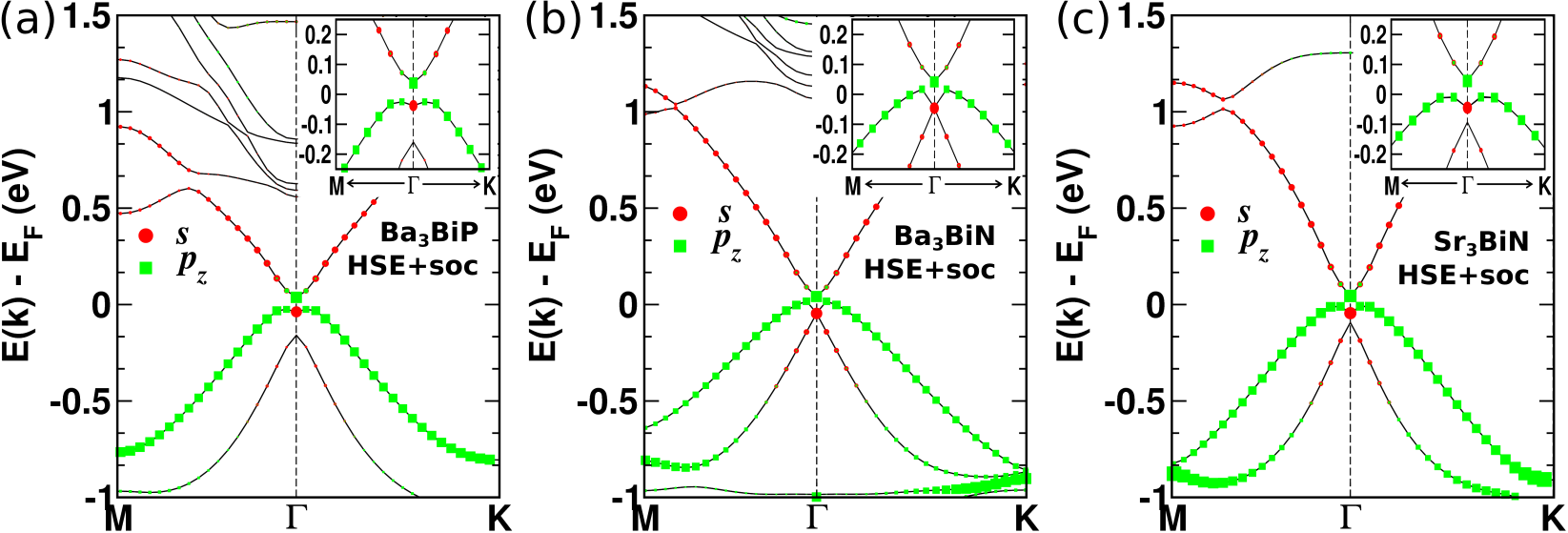}
\caption { (Color online) Hydrostatic pressure induced {\it s-p} band inversion for (a) Ba$_{3}$BiP, (b) Ba$_{3}$BiN  and (c) Sr$_{3}$BiN using HSE06+SOC. Size of red and green dots represent weightage contribution of Bi(Ba,Sr){\it -s} and N(P){\it -p$_{z}$} orbitals respectively. }
\label{fig3}
\end{figure}

{\par}{\it Ambient condition:}\ From the perspective of band inversion mechanism, we explore the possibility of realizing topological insulating phase in hexagonal antiperovskites. As mentioned before, we have presented results using HSE06 functional to get a more accurate picture of band gap and band inversion. However, for comparisons, we have used PBE functionals. Most of the results from PBE functionals are shown in supplement.\citep{supp} Table \ref{table-1} shows the theoretically relaxed lattice parameters for the three systems A$_3$BiB (A=Ba,Sr; B=P,N). The available experimental lattice parameters for Ba3BiN\cite{NIEWA2004} match fairly well with the simulated values. 

{\par} Figure \ref{fig2} shows the bulk band structure of three systems, at their theoretically relaxed lattice constants, using HSE06 functional including SOC.  Notably, all the three systems show trivial band ordering with a semi-conducting gap. The conduction band minima (CBM) and valence band maxima (VBM) are majorly contributed by $s$ and $p_z$-like states respectively. These bulk band structures were also simulated using PBE functional (see Fig. S1 of supplement\cite{supp}). Interestingly, the two systems, Ba$_3$BiP and Sr$_3$BiN are found to show band inversion at $\Gamma$-point between the extended $s$-like orbitals of (Ba,Sr)-Bi atoms and $p_z$-like orbitals of (P,N) atoms in the presence of SOC, while Ba$_3$BiN still holds trivial band ordering with a semi-conducting gap. Thus, the correct choice of exchange correlation functional is extremely important to make a reliable prediction for TI materials. This has been demonstrated in other studies as well, where PBE\cite{JOUBERT1999} functional is found to overestimate\citep{NaBaBi2016} the band inversion as compared to HSE06.\cite{HSE2003} 

{\par}{\it Under hydrostatic expansion:}\ Next, we have analyzed the HSE06 band structures under the application of tensile strain via hydrostatic expansion. With increasing lattice parameters  (hydrostatic expansion), the $s$-like orbitals shift downward and $p_z$-like states shift upward, eventually causing a band inversion between $s$ and $p_z$-like states at a critical volume. The electronic structures of Ba$_3$BiP, Ba$_3$BiN and Sr$_3$BiN at the critical expanded lattice parameter are shown in Fig.~\ref{fig3}. Within HSE06+SOC calculations,  Ba$_3$BiP, Ba$_3$BiN and Sr$_3$BiN show band inversion at a negative pressure of 1.22 GPa, 4.66 GPa, and 3.19 GPa respectively. The non-trivial band gap of Ba$_3$BiP and Sr$_3$BiN are found to be 61 meV  and 51 meV respectively. While for Ba$_3$BiN, the VBM just cross the Fermi level away from $\Gamma$-point and the band gap is negligibly small with a non-trivial band order. The HSE06 band structures in the absence of SOC show  trivial band order (shown in Fig.~S3 of supplement\citep{supp}).

{\par}  Band structures for the three compounds are also simulated using PBE functional, at their hydrostatically expanded lattice parameters. These are shown in Fig.~S4 of the supplement.\cite{supp} The non-trivial band gap in this case turns out to be  90 meV, 62 meV, and 84 meV for Ba$_3$BiP, Ba$_3$BiN and Sr$_3$BiN compounds respectively.

Since these systems posses inversion symmetry, the non-trivial band topology can be further confirmed by calculating the topological invariance {\it Z$_2$} index via parity analysis. Table~\ref{Table2} shows the parity eigenvalues of the occupied bands at time reversal invariant momenta (TRIM) points A, $\Gamma$, M and L . Product of all the parity eigenvalues yields {\it Z$_2$}=1 which again confirms the non-trivial band topology. All the calculations are performed including SOC effect.

\begin{table}[b]
\begin{ruledtabular}
\caption{Parity eigenvalues of occupied bands at TRIM points for the three systems in their non-trivial phase.   }
\label{Table2} 
\begin{tabular}{c c c c c c}
& A (0,0,$\pi$) &  $\Gamma$ (0,0,0) & 3M ($\pi$,0,0) & 3L ($\pi$,0,$\pi$) & product\\
\hline
Ba$_3$BiP & $-$ & $-$ & $+$ & $-$ & $-$ \\ 

Ba$_3$BiN & $-$ & $-$ & $+$ & $-$ & $-$ \\

Sr$_3$BiN & $-$ & $-$ & $+$ & $-$ & $-$ \\
\end{tabular} 
\end{ruledtabular}
\end{table}

\begin{figure}[t]
\centering
\includegraphics[width=\linewidth]{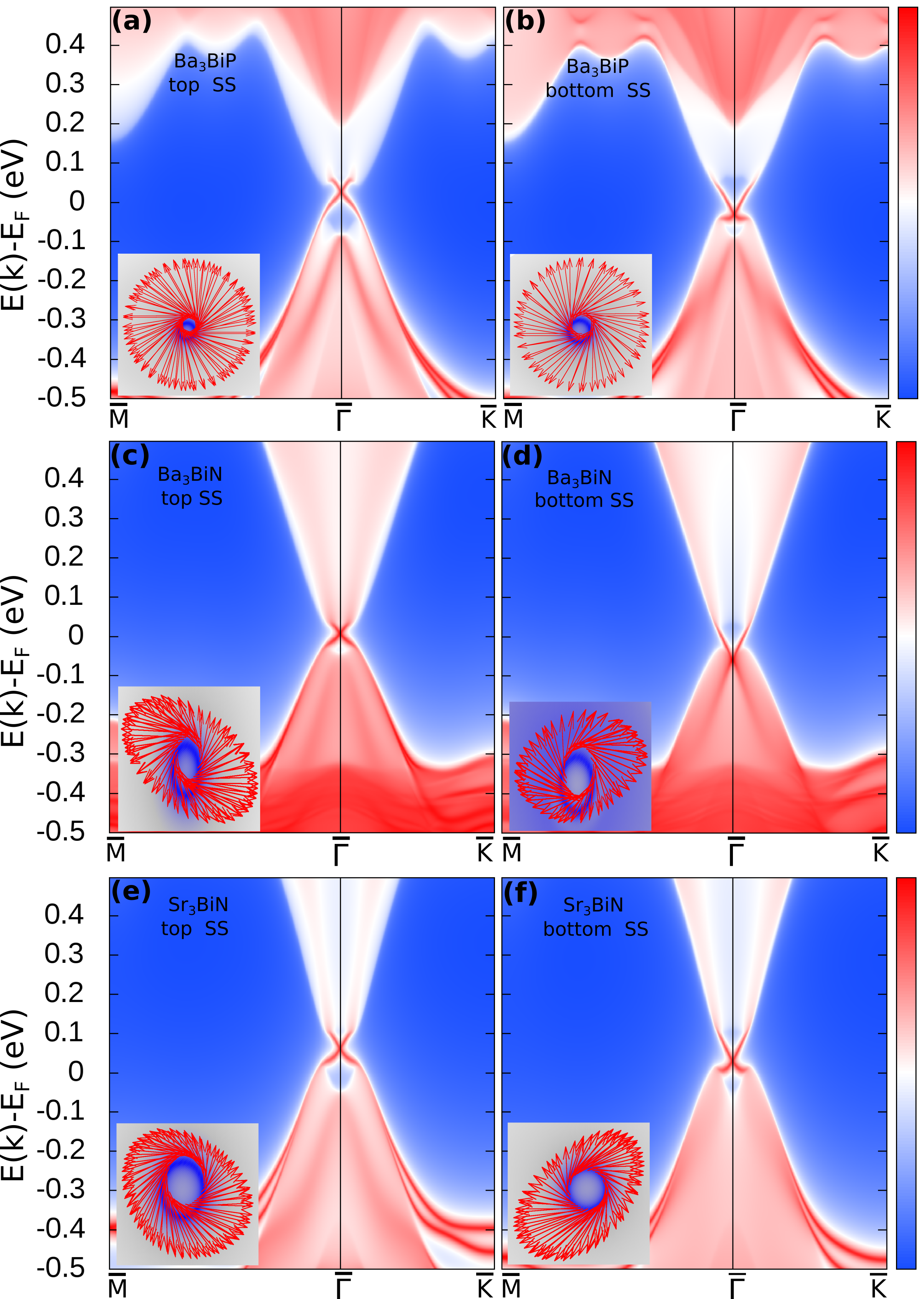}
\caption {(Color line) Projected surface density of states (SS) and their spin texture on (001) surface for Ba$_{3}$BiP, Ba$_{3}$BiN and Sr$_{3}$BiN. Left (Right) panels indicate the result for top (bottom) surfaces respectively. The spin texture of respective SSs are shown in the inset. The Dirac points (DP) of the top and bottom surface lie at different energy scale, indicating the asymmetry of the  potential at two surfaces. Red (blue) color in the spectral map represents highest (lowest) intensity. }
\label{fig4}
\end{figure}


{\it Topological surface states:}\ To further investigate the TI properties, we calculate the surface states on (001) surface using the slab model. We construct a tight binding (TB) Hamiltonian for the slab using MLWFs.\cite{MARZARI1997,SOUZA2001,VANDERBIT} It is important to note that, in general, the topology of the bands calculated from HSE06 and PBE functionals are similar in nature but with a different band gap (see Fig.~\ref{fig2} and Fig.~S1). Hence, we chose computationally less expensive PBE functional over HSE06 to construct the MLWFs for the three systems (but with HSE06 band gap value) in their non-trivial phase. Figure \ref{fig4} shows the surface spectral functions computed by the surface Green$^{'}$s function method\cite{QuanSheng,DHLEE_I,DHLEE_II,SANCHO1985} using TB Hamiltonian. Since the slab calculation involves two terminating surfaces, top and bottom, the corresponding projected surface density of states are shown in the left and right panels. The distinct feature of our proposed TIs is that the Fermi level lies in the non-trivial band gap and the Dirac-like linear surface dispersion exists at $\Gamma$-point inside the bulk gap (except the bottom surface Dirac cone of Ba$_3$BiN compound, which is buried inside the bulk states). Another benchmark of TIs are the spin momentum locking helical spin textures. To observe this, we have projected the spin directions in the slab Fermi Surface just above the Dirac point. The left (right) handed spin texture has been observed for top (bottom) surface as shown in the inset of Fig.~\ref{fig4}. Thus, the nature of surface states conclusively reconfirm the topological non-triviality of our three proposed TIs A$_3$BiB (A=Ba,Sr; B=P,N).

\begin{figure}[t]
\centering
\includegraphics[width=\linewidth]{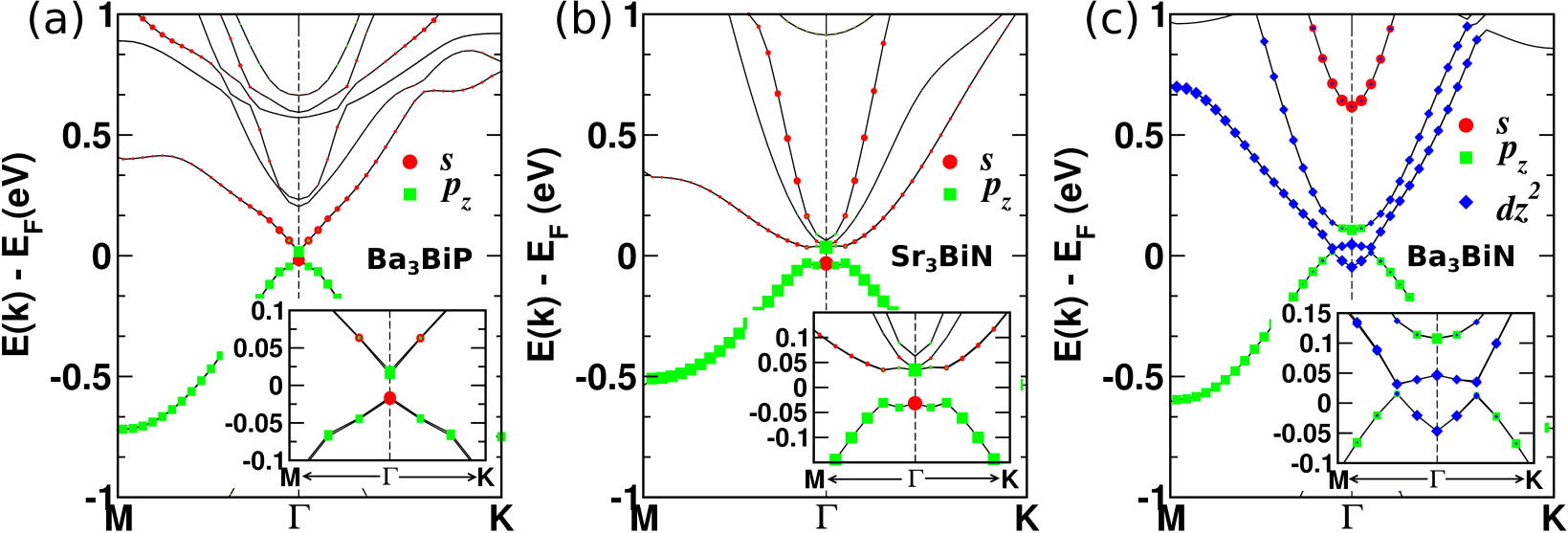}
\caption {(Color online) Bulk band structure of Ba$_3$BiP, Sr$_3$BiN and Ba$_3$BiN at uniaxial compressive pressure of 1.43 GPa, 13.75 GPa and 9.55 GPa within  HSE06+SOC calculation. Inset shows a closer look of non-trivial band topology around the $\Gamma$-point near the E$_F$. Red, green and blue symbols represent the orbital contribution of {\it s, p$_z$} and {\it dz$^2$}-states respectively.}
\label{fig5}
\end{figure}

{\par} {\it Under uniaxial compression:}\ Compressive strain is often easier to achieve in a given material. For this, we have checked the effect of compressive pressure along the {\it c}-axis of these systems. In the case of uniaxial compression, the orbitals aligned along {\it c}-axis play a crucial role in dictating the hybridization, and hence the band inversion. Figure \ref{fig5} shows the band structure of these three systems at/above a critical strain at which these systems show non-trivial band topology. Different colored symbols represent the weighted contributions of {\it s}, {\it p} and {\it d} orbitals. Inset shows a closer look of bands near E$_F$. A deeper analysis of the evolution of band topology shows that, as we increase the uniaxial compression in Ba$_3$BiN compound, the {\it d}-orbitals (which lie at a higher energy at the ambient conditions) start to come down and {\it s}-orbitals shift up. While the {\it p}-orbitals (mostly {\it p$_z$}) which initially lie in the VBM, start to shift up towards the Fermi level. At a critical compression, {\it p-d} inversion happens in Ba$_3$BiN as evident from Fig.~\ref{fig5}(c). Within the HSE06 calculation, this band inversion occurs at around 9.55 GPa, while in PBE it happens at a much lower pressure. The {\it d} bands which invert and move towards the valence band are majorly of {\it dz$^2$} character although there is a small contribution from {\it dx$^2$} and {\it dxy} as well. In contrast, for Ba$_3$BiP and Sr$_3$BiN compounds, with the enhancement of the compressive strain, both {\it d} and {\it s}-orbitals start to shift down towards the E$_F$. {\it s}-orbitals, however, move faster than {\it d} orbitals and gets inverted with the  {\it p$_z$} like states at $\Gamma$ point at a critical strain (see Fig.~\ref{fig5}(a,b)). The critical compressive pressure required for such a {\it s-p} inversion in Ba$_3$BiP and Sr$_3$BiN are 1.43 GPa and 13.75 GPa respectively, within the HSE06 calculation. Due to this band inversion, a non-trivial band gap opens up in Ba$_3$BiP and Sr$_3$BiN as evident from the inset of Fig.~\ref{fig5}(a,b). With increasing compressive pressure the amount of band gap increases keeping the non-trivial band order intact. However, for Ba$_3$BiN, the Fermi level closely crosses the valence band and the non-trivial band gap remains negligibly small (see the inset of Fig.~\ref{fig5}(c)). 

{\par} In order to investigate the possibility of experimental realization of these compounds in their hexagonal phase, we have calculated the formation energy and phonon dispersion which confirms the chemical and dynamical stability of these materials (see Sec. V of supplement\cite{supp}). 

{\par}{\it\bf  Model Effective Hamiltonian:}
To gain a better insight into the  nature of topological order around the $\Gamma$ point, we now use a low-energy {\bf k.p} simple model Hamiltonian around $\Gamma$ point. Such a Hamiltonian can be constructed using the method of invariants and considering the important crystal symmetries$-$ time-reversal, inversion, and D$_{6h}$ around $\Gamma $ point. Similar model Hamiltonian was also obtained for other topological materials.\cite{ZWANG2012,LiAuSe,Cd3As2}  As discussed earlier, the CBM and VBM are mainly contributed by p and s-like atomic orbitals, the corrosponding basis sets for the 4$\times$4 {\bf k.p} Hamiltonian within SOC effect can be taken as $|S^+_\frac{1}{2},\frac{1}{2}\rangle$, $|P^-_\frac{3}{2},\frac{3}{2}\rangle$, $|S^+_\frac{1}{2},-\frac{1}{2}\rangle$, $|P^-_\frac{3}{2},-\frac{3}{2}\rangle$. Using these basis sets, a minimal effective Hamiltonian around $\Gamma$ point for D$_{6h}$ point group is,

\[ H(\bf k) = \epsilon_{0}(\bf k) + \begin{pmatrix} 
M(\bf k) & A(\bf k)k_{+} & D(\bf k)k_{-} & -B^{*}(\bf k) \\
A(\bf k)k_{-} & -M(\bf k) & B^{*}(\bf k) & 0  \\
D(\bf k)k_{+} & B(\bf k) & M(\bf k) &  A(\bf k)k_{-}   \\
-B(\bf k) & 0 & A(\bf k)k_{+} & - M(\bf k)  \\

\end{pmatrix},  \]

where $ \epsilon_{0}({\bf k}) = C_0 + C_1k^{2}_{z} + C_2(k^{2}_{x}+k^{2}_{y}),\quad k_{\pm} = k_{x} \pm ik_{y},\quad A({\bf k}) = A_0 + A_{1}k^{2}_{z} + A_2(k^{2}_{x}+k^{2}_{y}),\quad M({\bf k}) = -M_0 + M_{1}k^{2}_{z} + M_2(k^{2}_{x}+k^{2}_{y})$ with $M_0$, $M_1$, $M_2$ $\textgreater$ 0 to ensure band inversion. Here, D({\bf k}) is the inversion symmetry breaking term, which is zero in our case. In such a situation, this Hamiltonian represents a pair of gapless 3D Dirac fermion states. The associated eigenvalues of the above 4$\times$4 Hamiltonian are, \\
\[ E({\bf k}) = \epsilon_{0}({\bf k})  \pm \sqrt{M({\bf k})^2 + A({\bf k})^2k_+k_- + |B({\bf k})|^2} \]

\begin{figure}[t]
\centering
\includegraphics[width=\linewidth]{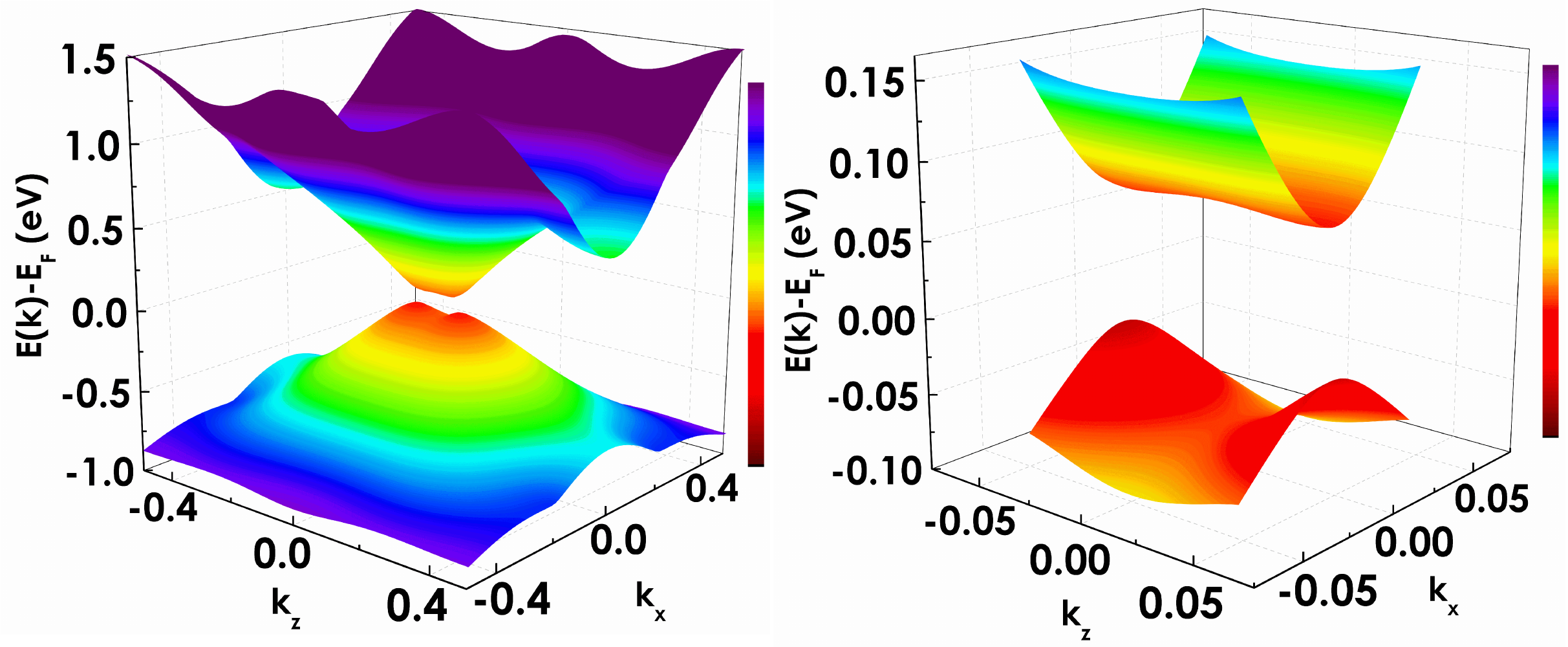}
\caption {(Color online) (Left) Ab-initio 3D band structure of Sr3BiN. (Right) Closer view of the 3D band where the model Hamiltonian is fitted.}
\label{fig6}
\end{figure}

Now, introducing a mass term in the above equation can yield a gapped state. This can be done by having  a linear leading order term in $ B({\bf k}) = B_1k_z $. We have fitted the energy spectrum of the above model Hamiltonian to the {\it ab-initio} PBE band structure for Sr$_3$BiN  by using a python based LMFIT package.\cite{lmfit}  The fitting parameters are $C_0$=0.0025, $C_1$=9.8962, $C_2$=8.1412, $M_0$=0.0727, $M_1$=12.5719, $M_2$=18.2233, $A_0$=0.6433, $A_1$=-49.7363, $A_2$=56.3998, and $B_1$=2.3345. Figure \ref{fig6} shows a  3D view of this band structure in two different wave vector range. The onset of band gap is clearly visible in the right panel where the model Hamiltonian is fitted.

{\par} In conclusion, we theoretically predict the possibility of realizing 3D TIs in the family of hexagonal antiperovskite systems.  To assess accurate band properties, we used HSE06 exchange correlation functional in our ab-initio calculation, which is known to predict more reliable band gap and inversion. At the ambient condition, all the three studied systems A$_3$BiB (A=Ba,Sr ; B=P,N) show trivial band ordering.  A topological phase transition is realized  by the application of  small amount of hydrostatic expansion. Non-triviality of the band order is further confirmed by calculating the topological {\it Z$_2$} index via parity analysis. Moreover, we have also simulated the surface spectral intensity maps using slab Green$^{'}$s function approach. Interestingly, we found Dirac-like surface states inside the bulk band gap. In addition to lattice expansion, non-trivial band ordering is also realized under uniaxial strain. Due to the different nature of hybridization, band inversion in this case occurs between the p- and d-orbitals as opposed to the conventional s- and p-orbitals.  The possibility of experimental realization of these compounds are checked and confirmed by simulating the chemical stability and phonon dispersion.

CKB acknowledges Indian Institute of Technology, Bombay for financial support in the form of teaching assistantship. AA also acknowledge fruitful scientific discussion with  Dr. Ambroise van Roekeghem.




\end{document}